\documentclass[a4paper,twocolumn,tightenlines,english,aps,prl,floatfix,showpacs]{revtex4}
\usepackage{babel}

\begin{document}

\title{Quantum Noise Minimization in Transistor Amplifiers}

\author{U. Gavish$^1$, B. Yurke$^2$ and Y. Imry$^3$}
\affiliation{1. Institute for Theoretical Physics, University of
Innsbruck, Innsbruck A-6020, Austria \\
2. Bell Laboratories, Lucent Technologies, Murray Hill, NJ
07974 \\
3. Condensed Matter Physics Dept., Weizmann Institute, Rehovot
76100, Israel}

\begin{abstract}
General quantum restrictions on the noise performance of linear
transistor amplifiers are used to identify the region in parameter
space where the quantum-limited performance is achievable and to
construct a practical procedure for approaching it  experimentally
using only the knowledge of directly measurable quantities: the
gain, (differential) conductance and the output noise. A specific
example of resonant barrier transistors is discussed.
\end{abstract}

\date{\today}

\pacs{42.50.Lc, 03.67.-a, 03.65.Ta, 73.23.-b}

\maketitle Heisenberg uncertainty relations restrict the
performance of amplifiers
 and detectors\cite{caves2}-\cite{aash}.
Derived from rather general properties (canonical commutation
relations for signals carried by non-conserved
bosons\cite{caves2}, or the nonequilibrium Kubo formula for other
signals\cite{Gavish gener constr}-\cite{aash}) such restrictions
specify the best-possible noise performance but do not provide a
procedure for obtaining it. For example, a (phase insensitive)
linear amplifier must add to the amplified signal a noise power of
at least $(G^2 - 1)\hbar\omega/2$ per unit bandwidth
\cite{impedance}, where $G^2$ is the power gain
\cite{caves2},\cite{Yurke denker},\cite{Gavish gener
constr},\cite{aash}. This restriction, referred to below as the
\emph{Heisenberg limit},
 is very general and applies e.g.
to laser amplifiers, parametric RF amplifiers, field effect
transistors, single electron transistors and molecular
transistors. However, the particular source of the noise varies
and therefore also the procedures one needs to follow in order to
minimize it. In parametric amplifiers this noise is the
\emph{equilibrium} current noise in the idler resistor\cite{Yurke
denker} and therefore this resistor should be cold enough to
produce only the zero point fluctuations.

In transistor devices, in which the amplification is performed by
a signal on a gate strongly modulating the output current, cooling
the device is not sufficient to obtain the ideal noise
performance. Such devices manifest \emph{nonequilibrium} noise
(called \emph{Idling-noise} below) in the source-drain current
even when the gate voltage is held fixed.
 When the gate is
connected to a signal source having nonzero impedance,
fluctuations in the gate potential will arise from fluctuations in
the number of charge carriers in the gate region. These gate
potential fluctuations cause additional source-drain current
fluctuations (called here \emph{amplified back-action} noise).

Using restrictions on the noise performance of (phase insensitive)
transistor amplifiers, we present a procedure for an experimental
identification of the region in parameter space where
quantum-limited noise performance is allowed (if such a region
exists). Constructed for practical purposes, this procedure only
makes use of the knowledge of quantities which are \emph{directly
measurable}. Neither a knowledge of the hamiltonian of the signal
source nor that of the transistor is required. As an example we
show how this procedure can achieve the Heisenberg limit in
certain resonant barrier transistors.

We begin by introducing the restrictions on the noise performance
of transistor amplifiers. Consider a signal carried by a current
$I_{in}$ which is flowing out of a source having a differential
conductance \cite{g} $g_s$ and which enters the amplifier input
port.  The resulting amplified signal $I_{out}$ is delivered to a
load resistor, having a differential conductance $g_{\ell}$,
connected to the amplifier output port. We shall consider an
amplifier which is impedance-matched to the load, i.e., it has an
impedance $g_{\ell}^{-1}$ at its output port. The constraints
presented below hold for this case. However, the noise
minimization procedure which is derived from them holds also in
the general case of impedance mismatch. If $I_{out}(t)$ is
proportional to $I_{in}(t)$ the amplifier is called linear (and
phase insensitive). One can then define the \emph{power gain},
$G^2,$ of the amplifier by the input-output relation $I_{out}(t) =
G \left(g_{\ell}/g_s\right)^{1/2} I_{in}(t).$  To be valid quantum
mechanically, this input-output relation must be augmented to have
the form
\begin{eqnarray}\label{1}
I_{out}(t) = G \sqrt{ \frac{g_{\ell}}{g_s} } I_{in}(t)+I_N(t)
\label{eq:iout}
\end{eqnarray}
where $I_{out}(t)=e^{iH_{tot}t}I_{out}(0) e^{-iH_{tot}t},$
$I_{in}(t)=e^{iH_{s}t}I_{in}(0) e^{-iH_{s}t},$ $H_{tot}=H_a+H_s+
\gamma H_{a,s}$ is the total hamiltonian, $H_s$ is the hamiltonian
of the signal source, $H_a$ is that of the amplifier, and $\gamma
H_{a,s}$ is that of the interaction between them. $\gamma$ is a
small dimensionless coupling constant.
 $I_N$ is called the noise current operator and is a function
of operators related to the amplifier degrees of freedom and
therefore commutes with $I_{in}:$ $[I_N(t),I_{in}(t)]=0.$ $I_N$ is
called
 'noise' because according to Eq.(\ref{1}) if the  source  is
prepared in an eigenstate of $I_{in}$ with an eigenvalue $i_{in},$
a single measurement of $I_{out}$ would yield the value $G\sqrt{
\frac{g_{\ell}}{g_s} }i_{in}$ \emph{plus} an additional
\emph{random} contribution from the amplifier, the fluctuations of
which are given by $\Delta I_N^2$ where $\Delta I^2\equiv \langle
I^2 \rangle -\langle   I \rangle^2$ (the average is taken with
respect to the amplifier state).

If the  signal source  and the amplifier are initially
 prepared in stationary states and if after switching on the coupling
 they remain in stationary states, although modified ones,
and if the amplifier remains approximately impedance matched, then
\cite{Gavish gener constr} $\Delta I_N^2\ge (G^2-1)
\frac{\hbar\omega_0}{2}g_{\ell} \Delta \nu $ where $ \Delta \nu
\equiv \Delta \omega /(2\pi)$ is the detection bandwidth and
$\Delta \omega $ is a narrow spread of frequencies around the
center frequency $\omega _0$ of the band in which the detection is
performed. This inequality is a constraint on the total amplifier
noise. \emph{Defining} the idling-noise current by $I_0\equiv
I_N(\gamma=0)$ and the amplified back-action noise current by $I_n
\equiv I_N(\gamma)-I_0$  and assuming these two contributions have
zero mean (for $\omega _0\neq 0$)
 and are uncorrelated, $\langle I_0I_n\rangle=0,$ one has $\Delta I_N^2=\Delta I_0^2+\Delta I_n^2,$
so that the above inequality restricts the sum of the two types of
noise. Assuming that $I_n\sim \gamma^2$ it is shown below that
their product is restricted by the condition \cite{molec amplif
moriond 2004},\cite{impedance}:
\begin{eqnarray}\label{3}
\Delta I_0(t)\Delta I_n(t)\ge G^2 \frac{\hbar\omega_0}{4}
g_{\ell}\Delta \nu
\end{eqnarray}
which implies that the Heisenberg limit for transistor amplifiers
with a large gain, $G^2\gg 1,$ is achieved if and only if
\begin{eqnarray}\label{4}
\Delta I_0^2=\Delta I_n^2=G^2\frac{\hbar\omega_0}{4}g_{\ell}
\Delta \nu
\end{eqnarray}
Eq.(\ref{3}) resembles  constraints derived  for general linear
detectors \cite{averin} and \cite{girvin} or specific ones
\cite{Averin2},\cite{aash}. It differs from these results in that
it contains only directly measurable quantities: the
 noise contributions one would measure  \emph{at the output}, the gain and the conductance.

Eq.(\ref{4}) has several nontrivial consequences. It shows that
the initial idling-noise  $\Delta I_0^2(t)$ should not be made too
\emph{small} since coupling a device with vanishing idling-noise
to a signal will result in the appearance of an amplified
back-action noise $\Delta I_n^2(t)$ which will diverge in order to
maintain the inequality in Eq.(\ref{3}). In particular, for ideal
operation of the amplifier at a given gain,
 the amplified back-action noise and the idling-noise
should be each equal to \emph{half} of the amplified zero point
fluctuations of the amplifier.

Before presenting a way to reach the condition Eq.(\ref{4}) in
practice, we outline the derivation of Eq.(\ref{3}) (for details
see Ref.  \cite{molec amplif moriond 2004}). Applying the
nonequilibrium Kubo formula\cite{kubo0}-\cite{Kubo g} to the
amplifier and the  source  one has:
\begin{eqnarray} \label{Kubo a}
\int_{-\infty}^{\infty} dt e^{i\omega t}\langle
[I_{\alpha}(t),I_{\alpha}(0)]\rangle=2\hbar \omega
g_{\alpha},~~~~~\alpha=a,s.
\end{eqnarray}
$g_{a}=g_{\ell}$ is the source-drain differential conductance of
the amplifier. $I_s$ is the  unperturbed current signal (i.e. the
source current in the   absence of coupling to the amplifier).
 $I_a$ is the current that would flow out of the amplifier
if the load resistor is replaced by a short \cite{Gavish gener
constr}. The impedance matching implies that $I_s=2I_{in}$ and
$I_a=2I_{out}.$
 Denoting $ I(\omega )=
\frac{1}{\sqrt{2\pi}}\int_{-\infty}^{\infty}d\omega I(t)e^{i\omega
t}, $ and $\bar{I}(\omega_0)\equiv \int_{\omega_0\pm
\frac{1}{2}\Delta \omega}I(\omega) e^{-i\omega t}d\omega$ and
using Eqs. (\ref{1}), (\ref{Kubo a}), and the fact that $I_{in}$
and $I_N$ commute, one has
\begin{eqnarray} \label{[IN,IN]}
\langle [\bar{I}_N(\omega_0),\bar{I}_N^{\dag}(\omega_0)]\rangle
=-(G ^2-1)\frac{\hbar\omega_0}{2}g_{\ell}\Delta \omega.
\end{eqnarray}
Subtracting Eq.(\ref{[IN,IN]}) written for $\gamma>0$ from itself
written for $\gamma=0$ and neglecting terms higher order than
$\gamma^2$ one obtains
$\langle[\bar{I}_n(\omega_0),\bar{I_0}^{\dagger}(\omega_0)]
+h.c.\rangle=-\pi G^2\hbar\omega _0 g_{\ell}\Delta \nu. $ Written
 as
 an expectation value of a commutator\cite{molec amplif moriond 2004},
$\langle
[\bar{I}_n(\omega_0)+\bar{I}_n^{\dag}(\omega_0),i(\bar{I}_0^{\dag}(\omega_0)-\bar{I}_0(\omega_0))]\rangle
=-i\pi G^2\hbar\omega_0g_{\ell}\Delta \nu,~ $ this leads to the
uncertainty relation Eq.(\ref{3}).

We now present a noise minimization procedure  aimed at obtaining
the two equalities in Eq.(\ref{4}) in devices in which the
Heisenberg limit is achievable.
 This procedure requires certain practical conditions
to hold, the main one being that the coupling $\gamma$ between the
 signal source  and the transistor gate can be
smoothly controlled over a wide range of values. It is also taken
for granted that the source-drain bias voltage $V$ is well
controlled. The control of the coupling can be achieved, for
example, by a control of the gate capacitance. The procedure
involves only the knowledge of measurable quantities - there is no
need to calculate in advance the $V$ and $\gamma$ dependence of
the noise. The procedure consists of two simple steps which we
refer to as \emph{noise balancing}  and \emph{gain matching.}
  In the first step, one varies the coupling and the bias voltage
until they reach two values, $\gamma_1$ and $V_1$ where the two
types of noise reach the same value:
\begin{eqnarray}\label{DIn=DI0 a}
\Delta I_{n}^{2}(V_1,\gamma_1)=\Delta I_{0}^{2}(V_1)
\end{eqnarray}
The functional dependence of the idling-noise on $V$ and $\gamma$
differs from that of the  amplified back-action noise (e.g.,
$I_0\sim \gamma^0$ while $I_n\sim \gamma^2$). Equating  the two
types of noise should therefore be possible by varying either
$\gamma$ or $V.$ The variation of both (and of other controllable
parameters)
 is in general necessary in order to maintain
the linearity of the amplifier.  The noise balancing does not
imply noise minimization and the total noise may even increase
during this step. In order to describe the step that follows noise
balancing, \emph{two} power gains are defined: The first, the
 \emph{signal} power gain $G^2(V_1,\gamma_1),$ is
determined by a direct gain measurement.  The second,  the
\emph{noise} power gain $G_{N}^2(V_1),$ is calculated using the
relation:
\begin{eqnarray}\label{7}
\Delta I_{0}^{2}(V_1)\equiv G_{N}^2(V_1)\frac{\hbar\omega _0}{4}
g_\ell \Delta \nu.
\end{eqnarray}
 $\frac{\hbar\omega _0}{4}
g_\ell \Delta \nu$ is half the power delivered by the zero point
fluctuations of the amplifier to the load. Therefore, $G_N^2$ is
the idling-noise referred to this power. The second step consists
of matching the two gains by varying the bias voltage and the
coupling until $G_{N}^2(V)= G^2(V,\gamma).$ This should be done
while maintaining the condition
\begin{eqnarray}\label{gmGp=const a}
\gamma G(\gamma,V)=const.
\end{eqnarray}
If  $G$ (as is often the case) $\sim\gamma V$, Eq.(\ref{gmGp=const
a}) means that the gain matching is performed while keeping the
product of $\gamma^2$ and the voltage constant: $\gamma^2
V=\gamma_1^2 V_1.$ Eq.(\ref{gmGp=const a}) ensures that the gain
matching is performed while keeping the idling-noise and amplified
back-action noise balanced as in Eq.(\ref{DIn=DI0 a}) and
therefore, the condition given by Eq.(\ref{4}) (and thus also the
Heisenberg limit) is achieved.

It remains to explain why the condition Eq.(\ref{gmGp=const a})
ensures that the two types of noise remain equal while the gains
are matched.
For this, we consider the origin of the amplified back-action
noise. Due to the linear coupling, a current fluctuation of order
$\Delta I_0$ in the transistor
 induces a fluctuation of order $\gamma \Delta I_0$ in the signal source.
This fluctuation is amplified and contributes a noise power $\sim
\gamma^2 G^2 \Delta I_0^2$ to the output signal. This extra noise
is the amplified back-action, $\Delta I_n^2.$ Thus,
\begin{eqnarray}\label{DIn/DI0=gmGp}
\frac{\Delta I_n^2}{\Delta I_0^2}\sim  \gamma^2 G^2,
\end{eqnarray}
which means that the ratio of the idling-noise and amplified
back-action noise remains constant if $\gamma^2 G^2$ does.

A typical example is where the idling noise is a shot-noise i.e.,
it results from the partitioning of charges between the two sides
of a tunnelling barrier in the source-drain current path. The
transfer of a fraction of this noise into the signal source stems
from transitions enabled by the appearance of new scattering
channels in the presence of the signal source where passing
electrons transfer a quantum of $\hbar\omega_0 $ to the signal
source. The total contribution of these processes is proportional
to the number of electrons in the transistor which can participate
in such transitions. At zero temperature, and if $\hbar\omega _0
\ll eV,$ all electrons in the nonequilibrium energy window created
by  the voltage $V$ may undergo such transitions and therefore the
number of these transitions is $\sim V.$ Thus, the power emitted
into the source  is $\sim \gamma^2 V.$ After amplification, the
contribution of these additional fluctuations in the signal
current, is $\Delta I_n^2\sim \gamma^2 V G^2.$ On the other hand,
the (low frequency) shot-noise power is \cite{shotnoise} $\Delta
I_0^{2,shot-noise} \sim V.$ These two estimates confirm
Eq.(\ref{DIn/DI0=gmGp}).

We now illustrate our results for the specific case of a signal
amplified by a resonant barrier transistor coupled capacitively to
a continuum of LC resonators (quantum harmonic oscillators) that
models a resistive signal source. The model is similar in many
features to those analyzed in Refs. \cite{Averin2}, \cite{Aash
RLM},\cite{Mozyrsky}.
\cite{Mozyrsky}. The total Hamiltonian is
\begin{eqnarray}  \label{Htot m}
H_{tot}=\sum_{i=1,2}\int_0^{\infty} d\epsilon \epsilon  b_i^{\dag}(\epsilon)b_i(\epsilon)+ \hbar \omega _AA^{\dag} A+  \nonumber\\
 +\sum_{i=1,2}
 \int_0^{\infty} d\epsilon\frac{ik(\epsilon)}{\sqrt{2\pi}} ( b_i^{\dag}(\epsilon) A -A^{\dag} b_i(\epsilon)  )\nonumber\\
  +\frac{A^{\dag} A e\hat{Q}_s}{C_g} +\int_B d\omega_s \hbar\omega_sa^{\dag}(\omega_s)a(\omega_s)~~~~~~~~~~ \end{eqnarray}
  where $\hat{Q}_s=\Delta Q(\omega_0)\int_B d\omega_s \frac{1}{\sqrt{\omega_s}}( a(\omega_s)+a^{\dag}(\omega_s) )$
 is the total charge on the capacitors in the LC oscillators and where
 $B=[\omega_0-\Delta\omega/2,\omega_0+\Delta\omega/2].$
The $b_i$'s, $A$'s and $a(\omega_s)$'s satisfy respectively
continuous fermionic, discrete fermionic and continuous bosonic
commutation relations. $b_i$ annihilates an electron in bath
$i=1,2$. $A$ annihilates an electron in the resonance level which
is located at energy $\hbar\omega_A$. $k(\epsilon)$ is the
tunnelling amplitude between the baths and the resonance level.
$\epsilon$ is the single-electron energy.  $k^2(\epsilon)$, which
is the resonance width, is taken to be wider than $eV$ so that the
\emph{second} derivative of the
 transmission with respect to $\epsilon,$ (but  not the first),
can be neglected. It is also assumed that $k^2(\epsilon)$ is small
compared to $\hbar\omega _A$ and the Fermi energy. $C_g$ is the
gate capacitance of the amplifier and $\Delta Q(\omega_0)$ is the
typical charge fluctuation in one of the oscillators in its ground
state, $\Delta Q=\sqrt{\hbar\omega_0 C/2}$ where $C$ is the
capacitance in each one of the LC circuits. Denoting the coupling
constant by $\gamma \equiv e\Delta Q/(C_gk^2)$ and assuming
$\gamma\ll 1,$
 the coupling term in $H_{tot}$ can be written as
 $A^{\dag} A e\hat{Q}_s/C_g=\gamma k^2 A^{\dag} A \hat{Q}_s/\Delta Q$
 which plays the role of $\gamma H_{a,s}$ above.
 The principle of operation of this transistor amplifier is the following:
 the signal modulates the position of the resonant level and hence the transmission.
 In the classical picture this modulates the output current. In the quantum picture,
 this creates inelastic
 components for the transmitted electrons which lead to a structure
 (proportional to the square of a large bias voltage)
mirroring the signal power spectrum in the output current power
spectrum.

The transistor is taken to be in a zero-temperature
 stationary state  with  bath 1 and 2  having chemical potentials
$\mu+eV$ and $\mu$ and thus occupation numbers  $
n_1(\epsilon)=\Theta (\epsilon)\Theta (\mu+eV-\epsilon)$ and
$n_2(\epsilon)=\Theta (\epsilon)\Theta (\mu-\epsilon).$ The
transistor current operator is defined by the rate of change in
the charge of the two baths:
\begin{eqnarray}\label{current operator m} I_{a}(t)=\frac{1}{2}(\dot{Q}_1(t)-\dot{Q}_2(t))\end{eqnarray}
 where $Q_i(t)=e\int_0^{\infty} d\epsilon
b^{\dag}_i(\epsilon,t)b_i(\epsilon,t)$ is the total charge in
 bath $i.$ Solving the Heisenberg equations of motion
to second order in $\gamma$ we find (recall: $I_{out}\equiv
\frac{1}{2}I_{a}$)
\begin{eqnarray} \label{Iout expanded m}
I_{out}(t)=I_0(t)+G
\sqrt{\frac{g_{\ell}}{\tilde{g}_s}}\tilde{I}_{in}(t)+I_n(t)+O(\gamma^3)
\end{eqnarray}
where $\tilde{I}_{in}\equiv \frac{1}{2}\omega _0Q_s(t)$ and
\begin{eqnarray} \label{Iow final m}
I_0(t)=\frac{e\hbar}{4\pi} \int_{\pm B} d\omega e^{-i\omega
t}\int_{-\infty}^{\infty}
d\omega'\times ~~~~~~~~~~~~~~~~~~~~\nonumber\\
(t^{\ast}(\omega') b_+ ^{\dagger}(\omega')
b_-(\omega'+\omega)+t(\omega') b_- ^{\dagger}(\omega'-\omega)
b_+(\omega')),
\end{eqnarray}
is the $\gamma=0$ current, $\pm B\equiv [-\omega _0-\Delta
\omega,-\omega _0+\Delta \omega]\cup [\omega _0-\Delta
\omega,\omega _0+\Delta \omega],$
 $ b_+=\frac{1}{\sqrt{2}}(b_1+b_2),$~$ b_-=\frac{1}{\sqrt{2}}(b_1-b_2),$
\begin{eqnarray} \label{14} G =\gamma\frac{eV}{\hbar\omega_0}T\sqrt{2(1-T)},\end{eqnarray}
$t(\omega)=-k^2/(-i\hbar(\omega -\omega _A)+k^2)$ is the
transmission amplitude at energy $\hbar\omega$, $T=|t|^2.$
$I_n(t)$ is the amplified back action noise current the explicit
expression for which will not be given here.
 Note that Eq.(\ref{Iout expanded m}) is an
\emph{operator} input-output relation and therefore enables one to
calculate expectation values of any function of $\tilde{I}_{in}$.
$g_{\ell}=Te^2/2\pi\hbar$ and $\tilde{g}_s\equiv \pi\Delta Q^2/
\hbar.$ $\tilde{g}_s$ is the differential linear response
 of the
 "current" $\tilde{I}_s=\omega_0 Q_s.$
Comparing Eqs.(\ref{1}) and (\ref{Iout expanded m}) one sees that
actually the device  performs linear amplification of
$\tilde{I}_s$ instead of $I_s=\dot{Q}_s.$  This is a consequence
of the capacitive coupling $H_{a,s}= e A^{\dagger } A Q_s/C_g .$
However, Eq.(\ref{Kubo a}) is valid also for $\tilde{I}_s$ and so
are all the above results - the only modification one needs to
apply is the replacement of $g_s$ by $\tilde{g}_s$ as done in
Eq.(\ref{Iout expanded m}).

Eq.(\ref{14}) implies that a large gain, $G ^2 \gg 1,$ requires a
stronger assumption than
  $eV\gg \hbar\omega_0$ namely, $eV \gg  \hbar\omega_0\gamma^{-1}.$
We also note that when solving the Heisenberg equations, the
coefficient  before $Q_s$ in Eq.(\ref{Iout expanded m}) turns out
to be  an \emph{operator}, $\hat{G}$ (Eq. 2, with $ G\rightarrow
\langle  \hat{G}\rangle$ is still valid in this case).  However,
for a narrow bandwidth signal, $\hbar\Delta\omega \ll eV,$ the
quantum fluctuations of this operator are negligible $\Delta
\hat{G}^2 \ll \langle \hat{G} \rangle^2\equiv G^2.$ This allows us
to replace it by its expectation value.

From Eqs.(\ref{Iout expanded m})-(\ref{14})  one obtains the
idling-noise:
\begin{eqnarray}\label{15}
\Delta I_0^2= T(1-T)\frac{e^3V}{4\pi\hbar}\Delta \nu,
\end{eqnarray}
A lengthier calculation yields the amplified back-action noise
\begin{eqnarray}\label{16}
\Delta I_n^2 = \frac{\gamma^4}{4}T^5(1-T)
(\frac{eV}{\hbar\omega_0})^2\frac{e^3V}{\pi\hbar}\Delta \nu.
\end{eqnarray}
One also finds that these noise sources are indeed uncorrelated
$\langle I_nI_0\rangle=0.$ Eqs. (\ref{14}), (\ref{15}) and
(\ref{16}) yield
\begin{eqnarray}\label{DI0DIn equal m}
\Delta I_0(t)\Delta I_n(t)= \frac{1}{4}G ^2
\hbar\omega_0g_{\ell}\Delta \nu,\end{eqnarray}
 Eqs. (\ref{15}), (\ref{16}) and (\ref{DI0DIn equal m}),
demonstrate how an amplifier satisfying the constraint
Eq.(\ref{3}) as  an equality, may still not be operating at the
Heisenberg limit.  To achieve this limit, the noise balancing
should be performed. Equating $\Delta I_0^2=\Delta I_n^2$
 yields the condition
 \begin{eqnarray}
\label{17} \gamma^2\frac{eV}{\hbar\omega _0}T^2=1.
 \end{eqnarray}
By Eqs. (\ref{4}), (\ref{7}), and (\ref{14})-(\ref{16}),  any pair
of $\gamma$ and $V$ satisfying Eq.(\ref{17}) results in
performance at the Heisenberg limit (i.e., here, $G_N=G$). By
Eq.(\ref{14}), Eq.(\ref{17}) implies that $\gamma G =const,$
confirming Eq.(\ref{DIn/DI0=gmGp}). To identify all possible
values for the gain at the Heisenberg limit, $G^{H},$ we insert
Eq.(\ref{17})
 into Eq.(\ref{14}) and find $G
^{(H)}=\sqrt{2(1-T)}/(\gamma T)$ .  One should recall the
assumption that the second derivative of the transmission vanishes
which is strictly true only when $T=3/4.$ Thus,
\begin{eqnarray}
 \label{g heisen}
  G ^{(H)}=\frac{2\sqrt{2}}{3}\frac{1}{\gamma}.
  \end{eqnarray}

To summarize, we presented a practical procedure for finding the
region in parameter space where transistor amplifiers achieve the
optimum noise performance allowed by quantum mechanics for linear
phase insensitive amplifiers. The procedure should be
experimentally feasible for linear devices for which such a
parameter region exists even if the precise hamiltonian
 of the device is unknown.  We then verified the validity of this procedure in the case of a  resonant
barrier transistor amplifier coupled to a resistive signal source
modelled as a continuum of LC resonators.

We are thankful for discussions with Y. Levinson. U. G. thanks A.
A. Clerk for commenting on the manuscript. Research at WIS was
supported by a Center of Excellence of the Israel Science
Foundation (ISF) and by the German Federal Ministry of Education
and Research (BMBF), within the framework of the German Israeli
Project Cooperation (DIP).


\begin{thebibliography}{99}


\bibitem{caves2} C. M. Caves, Phys. Rev. D \textbf{26}, 1817 (1982).

\bibitem{Yurke denker}
B. Yurke and J. S. Denker, Phys. Rev. \textbf{A 29}, 1419 (1984).

\bibitem{devoret set} M. H. Devoret
and R. J. Schoelkopf, Nature \textbf{406}, 1039 (2000) and
references therein.

\bibitem{averin} D. V. Averin, cond-mat/0301524.
\bibitem{girvin} A. A. Clerk, S. M. Girvin and A. D. Stone Phys. Rev. B \textbf{67}, 165324 (2003).

\bibitem{Gavish gener constr} U. Gavish, B. Yurke and Y. Imry,
Phys. Rev. Lett. 93, 250601 (2004).

\bibitem{molec amplif moriond 2004} U. Gavish et al.,
Proc. XXXIXth Rencontres de Moriond on Quantum Information and
Decoherence in Nanosystems, p. 73, Eds. D. C. Glattli et al., La
Thuille, January 2004. cond-mat/0407544.

\bibitem{Averin2} D. V. Averin, Fortschr. Phys. \textbf{48}, 1055, (2000).
\bibitem{Aash RLM} A. A. Clerk and A. D. Stone, Phys. Rev. \textbf{B} 69, 245303 (2004)

\bibitem{aash} A. A. Clerk, Phys. Rev. B 70, 245306 (2004).
Eq. (14) in that work can be shown to be consistent with
Eq.(\ref{3}) here when combining the assumptions in both works.


\bibitem{impedance}  This constraint is valid for a transistor which is impedance matched
to a load resistor connected to its output port which is what we
shall assume throughout this work. Although this assumption
affects the absolute value of the noise, it does not affect
\emph{the signal-to-noise ratio}. If, e.g., the load resistor is a
short (and is therefore no longer impedance-matched to the
transistor),
 both the noise and the signal powers are increased by a factor 4 with respect to their values
 in the impedance-matched setup, while the signal-to-noise ratio is unchanged.

\bibitem{g}$g_s(\omega) $ is
the linear response to a small AC field applied to a system in a
stationary state. If this state is an equilibrium, $g_s(\omega) $
is the ordinary conductance.

\bibitem{shotnoise} V.A. Khlus, JETP \textbf{66}, 1243 (1987);
G.B. Lesovik, JETP Lett., \textbf{49}, 592 (1989).



\bibitem{kubo0} R. Kubo, Can. J. Phys. 34, 1274 (1956).
\bibitem{Landau Lifshitz kubo}
 L. D. Landau and E. M. Lifshitz, \textit{Statistical Physics Part 1}, 3rd ed.,
Sec. 126, Butterworth Heinemann (1997).
\bibitem{Kubo g} See also Ref. \cite{Gavish gener constr} and references therein.

\bibitem{Mozyrsky}
D. Mozyrsky, I. Martin and M. B. Hastings, Phys. Rev. Lett 92,
018303 (2004).



\end{thebibliography}
\end{document}